\documentclass[pdflatex]{sn-jnl}
\usepackage{amsmath,amssymb,amsfonts}
\usepackage{float}
\usepackage{xcolor}
\newcommand{\highlight}[1]{\textcolor{black}{#1}}
\usepackage{tabularx,ragged2e,booktabs}
\newcolumntype{C}{>{\centering\arraybackslash}X}

\begin{document}


\title[MCFlash: Bulk Bitwise Processing in 3D NAND]{MCFlash: Bulk Bitwise Processing in 3D NAND with Dynamic Sensing and Multi-level Encoding}

\author*[1]{\fnm{Habib} \sur{Ur Rahman}}\email{habib.urrahman@colostate.edu}
\author[1]{\fnm{Tharini} \sur{Suresh}}\email{tharini@colostate.edu}
\author[1]{\fnm{Sudeep} \sur{Pasricha}}\email{sudeep.pasricha@colostate.edu}
\author[1]{\fnm{Biswajit} \sur{Ray}}\email{biswajit.ray@colostate.edu}

\affil[1]{
    \orgdiv{Electrical and Computer Engineering},
    \orgname{Colorado State University},
    \orgaddress{
        \city{Fort Collins},
        \state{CO},
        \postcode{80521},
        \country{USA}
    }
}


\abstract{This paper presents MCFlash, a practical and immediately deployable technique for executing bulk bitwise operations directly within commercial off-the-shelf (COTS) 3D NAND flash chips. MCFlash relies solely on standard user-mode instructions, combining Multi-Level Cell (MLC) data encodings with dynamically tuned read reference voltages to execute in-place bitwise operations. We evaluate MCFlash across diverse NAND flash chips, both floating-gate and charge-trap variants, \highlight{from different generations}. Our results represent the first demonstration of error-free, on-chip bitwise operations, sustaining over one billion operations on fresh blocks and maintaining bit-error rates below 0.015\% even after 10{,}000 program/erase (P/E) cycles.}


\maketitle


\section{Introduction}

Bulk bitwise operations are gaining prominence across a range of emerging application domains, including genome sequencing~\cite{ref1}, data analytics~\cite{ref2}, graph processing~\cite{ref3}, cryptography, and hyper-dimensional computing~\cite{ref4}. In traditional CPU/GPU-centric architectures, the performance and energy efficiency of these operations are often limited by the high cost of data movement between compute units and memory. Performing bitwise operations directly within raw NAND flash arrays, an approach known as in-flash computing (IFC), can drastically reduce data movement overhead and significantly enhance both performance and energy efficiency for data-intensive workloads.

Several prior works on IFC exploit the high bit density of NAND flash arrays to accelerate multiply-and-accumulate (MAC) operations for neural network applications~\cite{ref5}. These approaches typically rely on analog current accumulation, which requires substantial modifications to the NAND array structure and depends on expensive analog-to-digital converters. Another class of approaches, known as in-storage computing, utilizes the internal processors of \highlight{solid-state drives (SSDs)} or integrates hardware accelerators within SSD controllers to improve compute performance~\cite{ref6}. While in-storage computing can improve performance, it remains constrained by the data movement bottleneck between the raw NAND chips and the SSD controllers.

Two recent notable works on IFC for bulk bitwise operations are ParaBit~\cite{ref7} and Flash-Cosmos~\cite{ref8}. ParaBit exploits the control sequence of the latching circuitry in modern NAND flash chips to perform bitwise operations. It also leverages the cache read command, traditionally used to improve read performance by enabling data transfer from the NAND array to the flash controller in parallel with array sensing operations. To support this, NAND chips include a cache latch in addition to the standard sense latch. ParaBit implements bitwise logic by carefully orchestrating the activation sequence of these latch circuits. Although ParaBit requires no hardware modifications to the NAND array, it does necessitate changes to the timing control logic, which means a privileged command sequence must be developed for its execution. These changes are not straightforward in today’s NAND chips. Moreover, ParaBit supports only two-operand bitwise operations and is suited primarily for highly error-tolerant applications, as it does not incorporate mechanisms to mitigate the inherent error susceptibility of NAND flash memory.

Flash-Cosmos addresses the limitations of ParaBit by introducing two key innovations: Multi-Wordline Sensing (MWS) and Enhanced SLC-mode Programming (ESP). MWS enables bitwise operations on multiple operands in a single sensing cycle, significantly boosting throughput and performance. Meanwhile, ESP mitigates the error susceptibility of Single-Level Cell (SLC) NAND flash by increasing the voltage margin between program and erase states, thereby enabling zero-bit-error operation. However, implementing MWS incurs higher power and energy consumption, particularly in inter-block configurations where multiple blocks must be activated simultaneously, an uncommon practice in current NAND flash chip designs. Furthermore, realizing MWS requires a more complex wordline decoder circuitry, since existing NAND chips typically support only single-wordline sensing. Similarly, ESP demands modifications to the incremental step pulse programming (ISPP) algorithm, which is finely tuned for each NAND generation. ESP also increases latency, reduces density, and accelerates flash cell wear-out.

While ParaBit and Flash-Cosmos offer promising strategies for enabling in-flash bulk bitwise operations, they depend on specialized command sequences or require modifications to the NAND array architecture and biasing conditions. These dependencies limit their applicability on COTS NAND flash chips, as they typically cannot be implemented without direct support from the chip vendor. In contrast, our proposed solution in this work, called MCFlash (\underline{M}LC \underline{C}ompute \underline{F}lash), enables immediate deployment on COTS devices using only user-mode commands. It performs bulk bitwise operations by dynamically adjusting read reference voltages in MLC NAND and leveraging the logical data encoding of multilevel cells. Unlike Flash-Cosmos, MCFlash does not rely on MWS or modifications to the ISPP algorithm, avoiding the associated complexity, power overhead, and design modification concerns. The major contributions of this paper are as follows:
\begin{enumerate}
    \item We introduce MCFlash, a novel in-flash bulk bitwise operation technique that can be implemented on unmodified COTS NAND flash memory chips using existing user-mode commands.
    \item We demonstrate MCFlash with zero raw bit error rate (RBER) on \highlight{three different generations} of 3D NAND chips with floating-gate and charge-trap technologies from a major vendor.
    \item We provide guidelines to improve MCFlash in future generations of 3D NAND chips.
    \item We perform application-level analyses for image segmentation, encryption, and bitmap index workloads, and demonstrate that the MCFlash configuration achieves the best performance, up to 1.7$\times$ faster than ParaBit and over 16.5$\times$ and 12.6$\times$ faster than OSC and the ISC accelerator, respectively.
\end{enumerate}


\section{Background}\label{sec:bg}

\subsection{3D NAND flash memory based \highlight{SSDs}}

\begin{figure}[t]
\centering
\includegraphics[width=\linewidth]{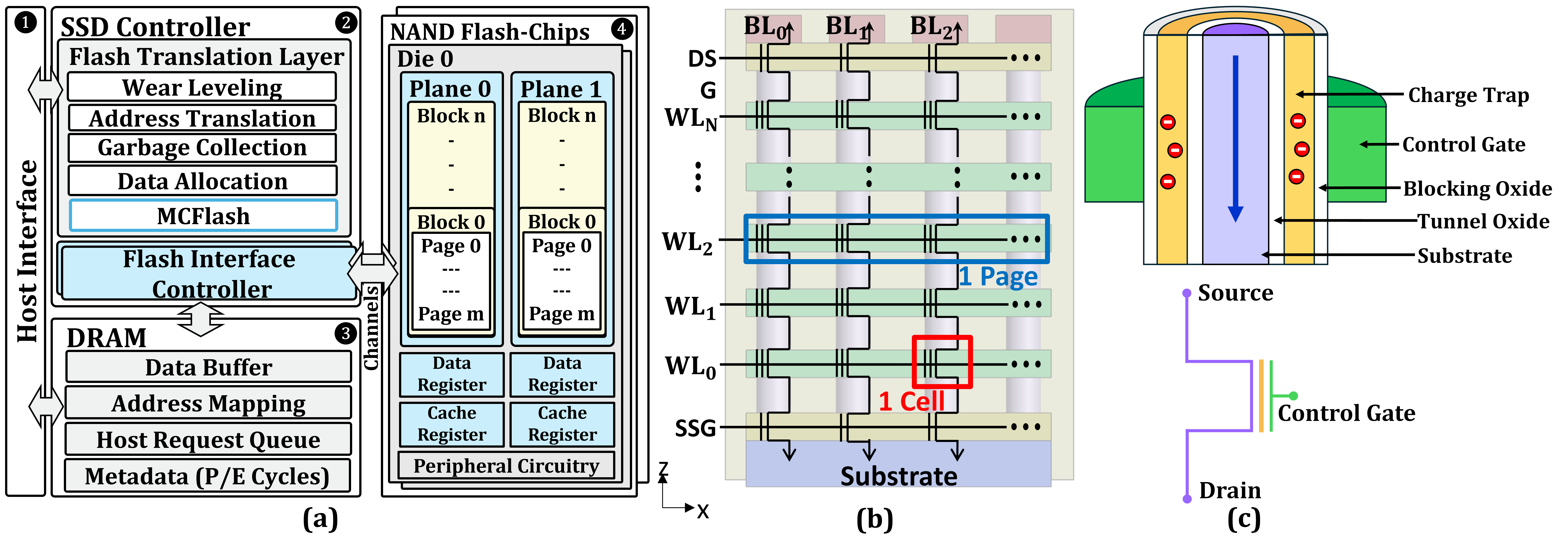}
\caption{3D NAND-based SSD architecture (a) System view of a SSD (b) Cross-section of a 3D NAND flash block (sub-block) (c) Cross-section and circuit symbol of a 3D NAND cell.}
\label{fig:1}
\end{figure}

Modern \highlight{SSDs} comprise four principal components as shown in Fig.~\ref{fig:1}: (1) Host interface, (2) SSD controller, (3) On-board DRAM, and (4) NAND flash chips. The host interface bridges the SSD and the host system, accepting I/O requests and returning data.

Internally, the SSD controller is a collection of embedded cores that implements the flash translation layer (FTL) and the multi-channel flash interface controller. The FTL encapsulates three critical functions: (a) address translation, which maps host logical block addresses to physical flash addresses; (b) wear leveling, which distributes erases and program cycles evenly across all flash blocks to maximize device lifetime; and (c) garbage collection, which identifies blocks containing invalid pages and reclaims them when the pool of free blocks falls below a threshold. Once the FTL issues a flash command, the flash interface controller generates the corresponding low-level command and address sequences and orchestrates data transfers.

Adjacent to the controller, DRAM serves three roles: it buffers host write data, caches the logical-to-physical (L2P) mapping table, and stores various metadata (e.g., P/E cycle counts or $N_{\mathrm{PE}}$). In addition, within an SSD, the NAND flash chips are arranged across multiple channels to exploit parallelism and increase internal data transfer bandwidth. Each NAND flash chip (Fig.~\ref{fig:1}) contains several independent dies, all sharing the same command/data channels. Within each die, two or more planes share peripheral circuitry. A plane comprises thousands of blocks, which are the granularity at which erase operations occur. Each block, in turn, contains thousands of logical pages (typically 8--16~kB each), which are the atomic units of read and program operations.

A closer view of the cross-section of the 3D NAND flash block structure is shown in Fig.~\ref{fig:1}(b), which is also referred to as a sub-block. In general, a 3D NAND block is physically divided into several identical sub-blocks to optimize die area utilization. All purple pillars within a sub-block represent one NAND string that connects to the substrate (bottom) when the corresponding source select gates (SSGs) are active, and to individual bit lines (BLs) at the top when the corresponding drain select gates (DSGs) are active. Each horizontal wordline connects one page’s worth of memory cells. In a 3D NAND array, the wordlines are vertically stacked as metal layers along the z-axis, with insulating dielectric layers separating them to form a complete memory block. Select transistors are used to select a particular sub-block during page read and write operations. However, all sub-blocks within a block are simultaneously erased during a block erase operation.

A single 3D NAND memory cell is highlighted with a red square in Fig.~\ref{fig:1}(b) and detailed in Fig.~\ref{fig:1}(c). Fig.~\ref{fig:1}(c) presents the cross-section of a single 3D NAND cell as a gate-all-around (GAA) transistor with a charge-trap layer embedded within the gate stack. The corresponding circuit symbol for a memory cell is also shown in Fig.~\ref{fig:1}(c). From inside out, the channel (purple) is surrounded by tunnel oxide, a charge-trap layer (yellow), blocking oxide, and the control gate (green). Data are encoded by the amount of charge stored in the CT layer: injecting electrons raises the cell’s threshold voltage ($V_{\mathrm{th}}$), representing a programmed “0,” while removing charge lowers the cell $V_{\mathrm{th}}$ to the erased “1” state in SLC technology.

\subsection{\texorpdfstring{Cell $V_{\mathrm{th}}$ distribution and logical encoding}{Cell Vth distribution and logical encoding}}

\highlight{ MLC NAND flash technology stores two bits per cell by controlling the cell's threshold voltage ($V_{\mathrm{th}}$) across four discrete levels. Thus, two logical pages share the same set of physical memory cells connected to the same wordline in MLC storage. The shared pages are named based on the significance level of their bits across the wordline. For example, all the least significant bits (LSB) of a wordline form the logical LSB page, while the most significant bits (MSB) of a wordline form the logical MSB page.}

\highlight{
Fig.~2 shows how MLC cell threshold voltages map to logical values. Since a memory row may contain up to 16~kB of cells, the analog $V_{\mathrm{th}}$ values are represented as $V_{\mathrm{th}}$ distributions ($L_0$--$L_3$ for MLC). Typically, Gray encoding is used for the $V_{\mathrm{th}}$ states to minimize bit errors by ensuring that consecutive states differ by only one bit. Given the user data corresponding to the shared LSB and MSB pages, the NAND controller determines the appropriate $V_{\mathrm{th}}$ states and performs the page program operation using the ISPP scheme.}

\highlight{
The decoding of $V_{\mathrm{th}}$ states occurs during a page read operation, which relies on sensing the cell's $V_{\mathrm{th}}$ using a set of \highlight{read reference} voltages ($V_{\mathrm{REF}}$). For instance, reading the LSB page of MLC memory requires only a single reference voltage $V_{\mathrm{REF1}}$ (blue dashed line in Fig.~2). All cells with $V_{\mathrm{th}}<V_{\mathrm{REF1}}$ are interpreted as logical ‘1’, while those above are interpreted as logical ‘0’. In contrast, reading the MSB page is more complex and requires two reference voltages, illustrated by red vertical lines in Fig.~2. First, $V_{\mathrm{REF0}}$ is applied on the wordline, and cells with $V_{\mathrm{th}}<V_{\mathrm{REF0}}$ are decoded as logical ‘1’. Next, $V_{\mathrm{REF2}}$ is applied, and cells with $V_{\mathrm{REF0}}<V_{\mathrm{th}}<V_{\mathrm{REF2}}$ are interpreted as logical ‘0’, while cells with $V_{\mathrm{th}}>V_{\mathrm{REF2}}$ represent logical ‘1’.}

\begin{figure}[t]
\centering
\includegraphics[width=\linewidth]{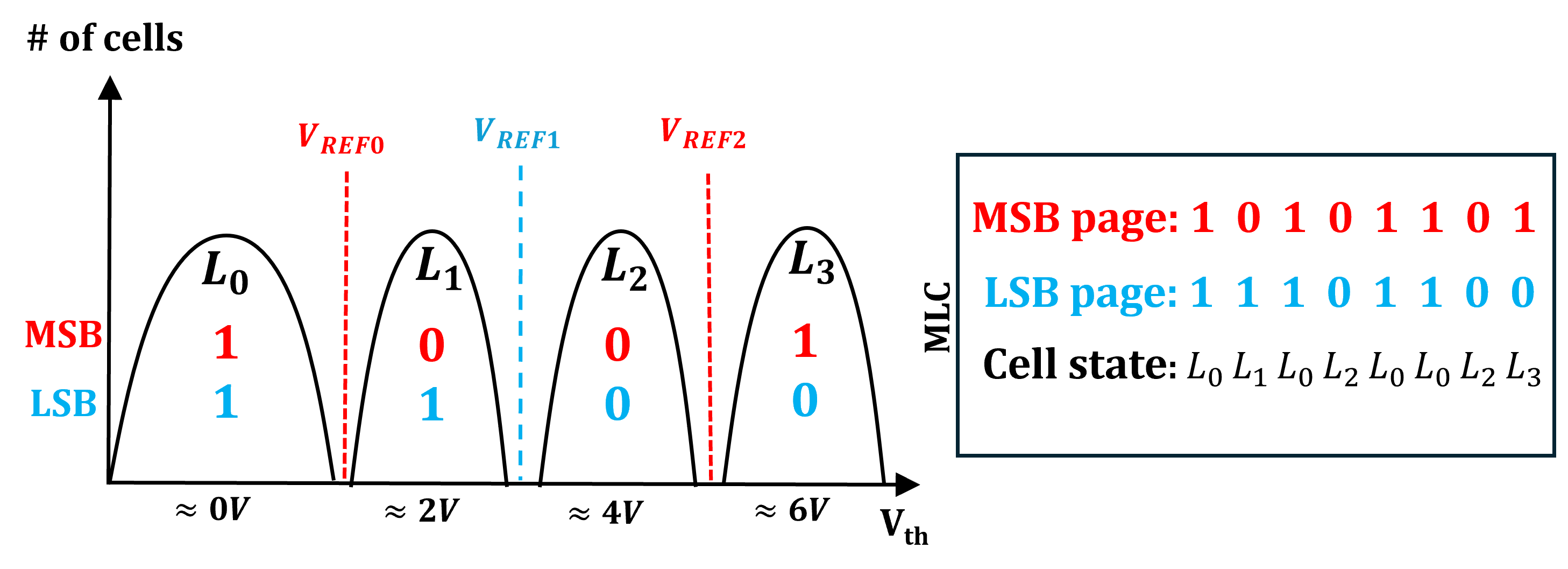}
\caption{MLC cell threshold voltage distributions and logical encoding. The inset illustrates the mapping between logical LSB/MSB page bits and physical cell states.}
\label{fig:2}
\end{figure}


\section{Related Work}\label{sec:relatedworks}

Data movement between memory/storage modules and the processor remains a critical performance bottleneck in modern compute-centric systems. To address this challenge, recent research has explored novel computing paradigms that enable near-data processing (NDP). In the following, we describe recent progress on three distinct NDP paradigms: in-memory computing (IMC), in-storage computing (ISC), and in-flash computing.

\textbf{In-flash computing:} IFC is a computing paradigm in which data processing occurs directly within NAND flash chips. By enabling computation inside the storage medium itself, IFC significantly reduces data movement overhead across the memory hierarchy. Prior research on IFC has primarily focused on executing multiply-and-accumulate (MAC) operations within the flash array by leveraging its inherent analog characteristics \cite{ref9,ref10,ref11,ref12,ref13}. For example, these works exploit analog current accumulation in the NAND string for implementing neural networks and mixed-signal applications with substantial energy and latency benefits. However, such implementations often require significant modifications to the NAND architecture, including integrating high-precision and costly analog-to-digital converters (ADCs) within the chip. As a result, these designs are not readily applicable to commodity NAND flash technologies.

In contrast to analog IFC, recent studies \cite{ref7,ref8} have explored digital computation within NAND flash arrays, enabling bulk bitwise operations without requiring substantial hardware modifications to commodity NAND chips. ParaBit \cite{ref7} performs in-place bitwise logic operations by carefully controlling the activation sequence of latching and sensing circuits, leveraging the existing architecture of modern NAND flash without physical changes. Similarly, Flash-Cosmos \cite{ref8} introduces MWS and ESP to execute bitwise operations across multiple operands within a single sensing cycle. These approaches achieve high throughput with zero-bit error rates, demonstrating the practicality of digital IFC on unmodified NAND arrays. MCFlash extends this line of work by enabling deployable bulk bitwise operations on COTS NAND chips using only user-mode commands.

\textbf{In-storage computing:} ISC architectures offload computation to dedicated processing units embedded within storage devices. Prior work has explored various ISC implementations, including drive-integrated ARM processors \cite{ref14,ref15,ref16,ref17} and custom hardware accelerators \cite{ref18,ref19,ref20,ref21,ref22}. These designs reduce data movement by transferring only the final results to the host system, improving performance and energy efficiency. However, ISC architectures face practical limitations: embedded CPUs typically offer limited compute power, accelerator-based solutions often require significant programming effort and lack flexibility, and most importantly, ISC still depends on internal data transfers between flash memory and the SSD controller, introducing latency and remaining a fundamental bottleneck.

\textbf{In-memory computing:} IMC has gained significant attention as a promising computing paradigm that mitigates the data movement bottleneck by enabling computation directly within main memory arrays \cite{ref23,ref24,ref25,ref26,ref27,ref28,ref29,ref30,ref31,ref32,ref33,ref34} and SRAM caches \cite{ref35,ref36,ref37,ref38}. A representative example is Ambit \cite{ref25}, which utilizes intrinsic DRAM properties such as charge sharing to perform in-memory computations. While IMC reduces latency and improves energy efficiency, its effectiveness is limited by the size of on-chip caches and main memory. When dataset sizes exceed memory capacity, substantial data transfer between memory and storage becomes unavoidable, reintroducing the same bottleneck.

\textbf{Pinatubo:} Pinatubo proposes bulk bitwise operations within non-volatile memory arrays, but MCFlash differs from Pinatubo in key aspects. First, Pinatubo relies on multi-wordline (multi-row) sensing for bitwise operations, whereas MCFlash uses single-wordline sensing (standard read operations). Second, Pinatubo is designed for SLC memories, whereas MCFlash operates on MLC NAND flash. Third, Pinatubo often requires modifications to peripheral circuits such as sense amplifiers and write drivers, while MCFlash requires no hardware changes. Fourth, 3D NAND provides significantly higher bit-level parallelism, with typical row sizes up to 16~kB, far larger than conventional byte-addressable NVMs. Modern 3D NAND dies can integrate more than one terabit of data, enabling operand placement within the same die and minimizing inter-die data transfers.


\section{Design of MCFlash}\label{sec:design}

\subsection{Key mechanism}\label{sec:keymechanism}

\begin{figure}[b]
\centering
\includegraphics[width=\linewidth]{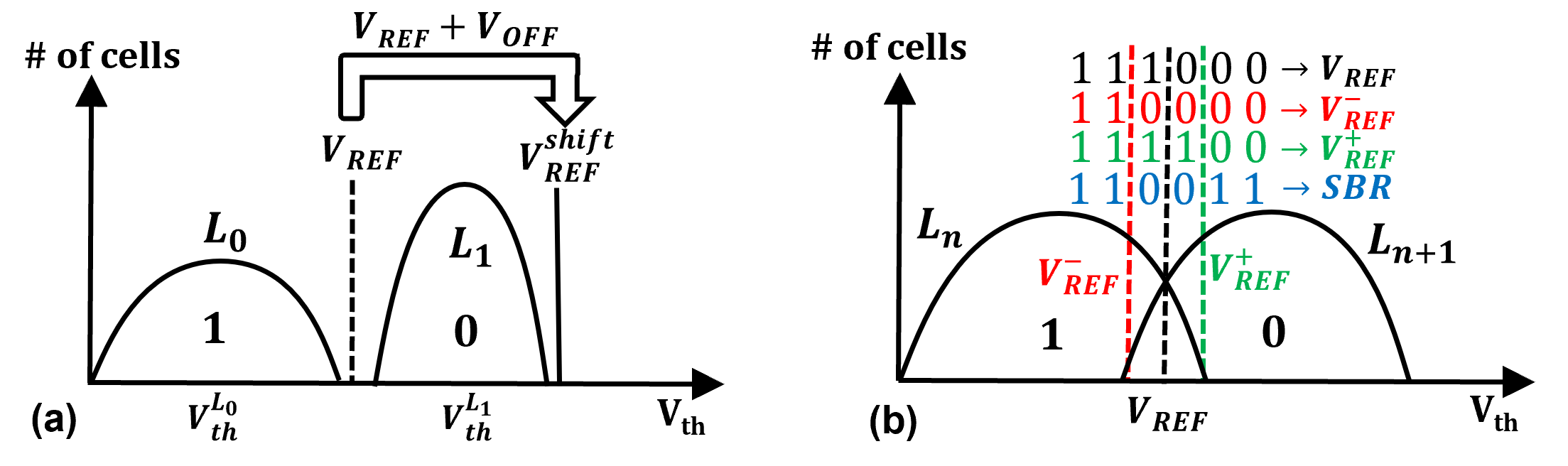}
\caption{Key mechanisms behind MCFlash operations. (a) \highlight{Shifted Read}: shifting the read reference voltage to realize different logic functions. (b) SBR: using two references and internal XNOR to generate soft bits.}
\label{fig:3}
\end{figure}

MCFlash leverages the inherent logical data encoding of MLC NAND flash to enable in-place bulk bitwise operations. It relies on two fundamental operations supported by raw NAND flash chips:

\highlight{\textbf{Shifted Read:}} The core mechanism of MCFlash involves dynamically adjusting the \highlight{read reference} voltage during sensing. This process is illustrated in Fig.~\ref{fig:3}(a), where the default reference voltage $V_{\mathrm{REF}}$ is marked by a dashed vertical line. A page read using the default $V_{\mathrm{REF}}$ correctly interprets cell states, reading cells in the $L_1$ state as ``0’’ and those in the $L_0$ state as ``1’’. However, COTS raw NAND chips provide user-accessible commands to shift this reference voltage so that the storage system controller can enhance data integrity by compensating for threshold voltage variations due to wear or data retention. This shift is controlled by a key design parameter called the read offset voltage $V_{\mathrm{OFF}}$, which can be either positive or negative. Thus, the shifted reference voltage is defined as $V_{\mathrm{REF}}^{\mathrm{shift}} = V_{\mathrm{REF}} + V_{\mathrm{OFF}}$. For instance, shifting the reference voltage to a new value $V_{\mathrm{REF}}^{\mathrm{shift}}$, as shown in the figure, may cause all cells, regardless of their programmed threshold voltage, to be read as ``1’’. By carefully selecting $V_{\mathrm{OFF}}$, MCFlash can perform specific bitwise operations during readout of MLC pages.

\textbf{Soft Bit Read (SBR):} The SBR is another feature offered by raw NAND chips to system integrators, aimed at improving data integrity during error correction. As illustrated in Fig.~\ref{fig:3}(b), SBR involves using two additional reference voltages $V_{\mathrm{REF}}^{+}$ (shown in green) and $V_{\mathrm{REF}}^{-}$ (in red), in addition to the default reference voltage $V_{\mathrm{REF}}$ (in black). A read using the default $V_{\mathrm{REF}}$ is typically referred to as a hard read. In contrast, SBR performs soft sensing by reading the same page twice: once with $V_{\mathrm{REF}}^{-}$ and once with $V_{\mathrm{REF}}^{+}$. The results of these two reads are then combined using an internal bitwise XNOR operation to generate a soft bit (in blue). The primary goal of SBR is to identify tail bits, which lie in the overlap region between adjacent threshold voltage distributions. By detecting these marginal cells, the error correction engine~\cite{ref39} can improve decoding accuracy and reduce the overall error rate. In this work, we carefully select the positions of $V_{\mathrm{REF}}^{+}$ and $V_{\mathrm{REF}}^{-}$ during SBR operations to achieve certain bulk bitwise operations.

Note that \highlight{Shifted Read} (or Read Retry) and SBR are standard features in NAND flash, typically supported through user-accessible commands~\cite{ref40}. While these commands are primarily designed to improve data integrity in the presence of threshold voltage variations, we repurpose them in this work to implement logical operations in COTS NAND flash.

\subsection{Bulk-bitwise operations using MCFlash}\label{sec:bulk-bitwise}

\begin{figure}[b]
\centering
\includegraphics[width=\linewidth]{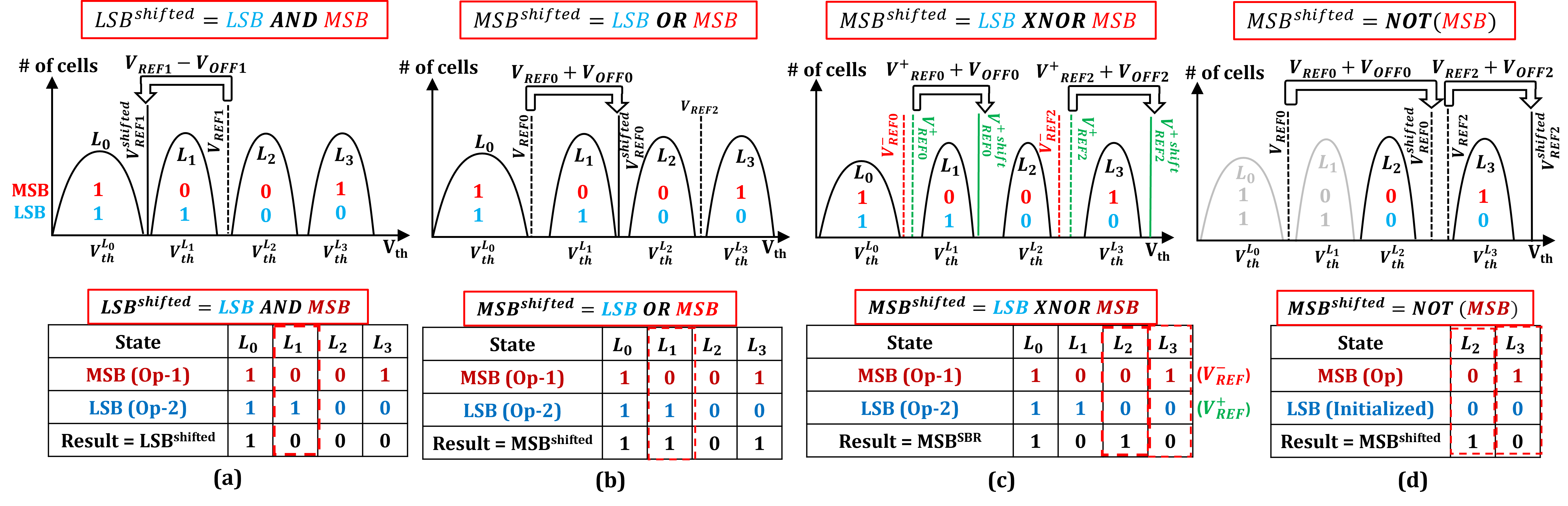}
\caption{Overview of MCFlash bulk bitwise operations. (a) AND, (b) OR, (c) XNOR, and (d) NOT with corresponding bit-level decoding based on shifted read and SBR mechanism.}
\label{fig:4}
\end{figure}

\textbf{Bitwise AND operation:} We illustrate the implementation of MCFlash in Fig.~\ref{fig:4}(a), starting with the bitwise AND operation. In this scheme, the two operands are stored on the shared LSB and MSB pages. A shifted read, as illustrated in Fig.~\ref{fig:4}(a), on the LSB page then directly produces the result of the bitwise AND. Note that the bitwise AND operation should yield the result $(1,0,0,0)$ for the cell states $(L_0,\, L_1,\, L_2,\, L_3)$ as shown in the table of Fig.~\ref{fig:4}(a). In other words, only cells in the $L_0$ state must be sensed as logic “1’’ during the AND operation, while all other states should be interpreted as logic “0’’. This is accomplished by lowering the reference voltage $V_{\mathrm{REF1}}$ below the $L_1$ threshold distribution by applying a negative offset voltage during the LSB page read, as shown in Fig.~\ref{fig:4}(a). Because NAND flash decoding depends on the relative position of a cell’s threshold voltage with respect to the \highlight{read reference} voltages, this shift causes the LSB bit of $L_1$-state cells to be interpreted as “0’’ although their original written value is “1’’.

\textbf{Bitwise OR operation:} The bitwise OR operation is illustrated in Fig.~\ref{fig:4}(b). The overall mechanism is similar to that of the AND operation; however, instead of reading the LSB page, the OR operation is implemented by reading the MSB page. Note that the bitwise OR operation should yield the result $(1,1,0,1)$ for the cell states $(L_0,\, L_1,\, L_2,\, L_3)$ as shown in the table of Fig.~\ref{fig:4}(b). Thus, cells in the $L_2$ state must be interpreted as “0”, while all other states should be read as “1”. This behavior can be achieved by appropriately shifting the read reference voltages during the MSB read, as shown in Fig.~\ref{fig:4}(b). Specifically, by upshifting $V_{\mathrm{REF0}}$ to a position between $L_1$ and $L_2$ states, similar to the default position of $V_{\mathrm{REF1}}$, the MSB page can be read in a way that effectively performs a bitwise OR operation. Note that the bitwise OR operation incurs higher latency compared to the AND operation, as reading the MSB page typically takes nearly twice as long as reading the LSB page. This is because the MSB read requires two sensing operations corresponding to two different reference voltages, whereas the LSB read involves only one.

\textbf{Bitwise XNOR operation:} To implement the bitwise XNOR operation, the read offset mechanism alone is not sufficient; instead, we must combine it with the SBR technique, as illustrated in Fig.~\ref{fig:4}(c). Recall that SBR inherently performs a bitwise XNOR between two reads one using positive sensing with $V_{\mathrm{REF}}^{+}$ and the other using negative sensing with $V_{\mathrm{REF}}^{-}$. We leverage this functionality to compute the XNOR of two operands stored in the LSB and MSB pages as follows:

\begin{enumerate}
    \item The negative sensing read is performed using the default reference voltages, producing the original MSB page data (operand~1).
    \item During the positive sensing phase, we apply read offsets such that the sensing operation effectively retrieves the LSB page data (operand~2).
\end{enumerate}

As shown in Fig.~\ref{fig:4}(c), the shifted read during the positive sensing phase aligns the sensing thresholds so that the MSB read mimics the LSB data, regardless of the original stored values. Since SBR outputs the XNOR of the two sensed values, this setup directly yields the bitwise XNOR result of the LSB and MSB operands.

In the following, we illustrate a step-by-step example of the XNOR operation. Note that the bitwise XNOR operation between LSB and MSB pages should yield the result $(1,0,1,0)$ for the cell states $(L_0,\, L_1,\, L_2,\, L_3)$. When the MSB page is read using negative sensing with default references, the internal page buffer stores $(1,0,0,1)$ for these states. In contrast, reading the same page with shifted read references during positive sensing (green lines in Fig.~\ref{fig:4}(c)) produces $(1,1,0,0)$, which mirrors the corresponding LSB page. The SBR operation then performs an internal bitwise XNOR between the two sensing results, yielding $(1,0,1,0)$ for $(L_0,\, L_1,\, L_2,\, L_3)$, which is the expected output.

\textbf{Bitwise NOT operation:} 
Since NOT is a unary bitwise operation, only one page is required to store the operand. However, in MLC storage, the corresponding shared page must be initialized with a specific data pattern prior to the write operation. In our approach, the operand is stored in the MSB page, while the LSB page is initialized with an all-zero data pattern. The choice of an all-zero initialization rather than all-ones is due to the characteristics of the erase distribution ($L_0$), which is typically wider. If the LSB were initialized to all-ones, the NOT operation on the MSB page would require shifting the read reference voltage $V_{\mathrm{REF0}}$ below the broad $L_0$ distribution. This level of downward shift is generally not achievable within the user-accessible read offset range. The NOT implementation is illustrated in Fig.~\ref{fig:4}(d) with the LSB page initialized with all-zero. This initialization eliminates the $L_0$ and $L_1$ states, and MSB data is written through the $L_2$ and $L_3$ states. To implement the NOT operation on the MSB page, the $L_2$ state needs to be read as “1’’ and the $L_3$ state as “0’’. This can be achieved by shifting both $V_{\mathrm{REF0}}$ and $V_{\mathrm{REF2}}$ as shown in Fig.~\ref{fig:4}(d).

\textbf{Inverse read and other bitwise operations:} Inverse read is another feature that certain raw NAND chips offer to system integrators to read the inverse value of the stored data~\cite{ref41}. Inverse read naturally provides a bitwise NOT operation without involving any page write operation. Additionally, other bitwise operations such as NAND, NOR and XOR can be implemented using inverse read along with the procedures for AND, OR and XNOR operations as described previously. Table~\ref{tab:mcflash-offsets} provides a concise overview of the read offset values essential for performing various bitwise operations with MCFlash. Each required read offset value is indicated as $\pm\Delta V_{\mathrm{th}}^{L_n}$, representing the necessary voltage shift to cross state $L_n$. The sign indicates the direction of the shift: a positive sign indicates an upward (forward) shift and a negative sign indicates a downward (backward) shift relative to the default reference voltage position. For instance, as illustrated in the bitwise OR column of MCFlash, a $+\Delta V_{\mathrm{th}}^{L_1}$ shift is applied to $V_{\mathrm{REF0}}$, while the other reference remains at its default position. This configuration is consistent with the practical implementation demonstrated in Fig.~\ref{fig:4}(b) for the bitwise OR operation.

Note that NOR, NAND and XOR operations can also be implemented without invoking inverse read, as summarized in Table~\ref{tab:mcflash-offsets}. However, the challenge in implementing them is the downward adjustment of the reference voltage $V_{\mathrm{REF0}}$ below the erased-state threshold ($L_0$). Since the erase distribution exhibits a substantially wider spread than the programmed distributions, and the standard read-offset values are optimized for mitigating errors within the programmed-state window ($L_1$--$L_3$), relying solely on user-command-based read offset adjustments may fail to fully traverse the erased-state window. This incomplete shift may lead to elevated raw bit error rates when implementing NOR, NAND and XOR without the inverse read technique.


\setlength{\textfloatsep}{8pt}

\begin{table}[t]

\caption{\highlight{Summary of read offsets for MCFlash operations.}}
\label{tab:mcflash-offsets}
\centering
\renewcommand{\arraystretch}{2.4}
\newcolumntype{Y}{>{\centering\arraybackslash}p{0.35cm}}
\tiny
\begin{tabularx}{\textwidth}{@{} c C Y C C Y C C C Y C @{}}
\toprule
 & \textbf{NOT} & \textbf{AND} & \textbf{OR} &
 \multicolumn{2}{c}{\textbf{XNOR}} &
 \textbf{NAND} &
 \multicolumn{2}{c}{\textbf{NOR}} &
 \multicolumn{2}{c}{\textbf{XOR}} \\
\cmidrule(lr){5-6}\cmidrule(lr){8-9}\cmidrule(lr){10-11}
$\mathbf{V_{\mathrm{REF}}}$ & & & &
 $\mathbf{V_{\mathrm{REF}}^{+}}$ & $\mathbf{V_{\mathrm{REF}}^{-}}$ &
 & $\mathbf{V_{\mathrm{REF}}^{+}}$ & $\mathbf{V_{\mathrm{REF}}^{-}}$ &
 $\mathbf{V_{\mathrm{REF}}^{+}}$ & $\mathbf{V_{\mathrm{REF}}^{-}}$ \\
\midrule
$\mathbf{V_{\mathrm{REF0}}}$ &
 $(\Delta V_{\mathrm{th}}^{L_1} + \Delta V_{\mathrm{th}}^{L_2})$ &
 $0$ &
 $+\Delta V_{\mathrm{th}}^{L_1}$ &
 $+\Delta V_{\mathrm{th}}^{L_1}$ &
 $0$ &
 $-\Delta V_{\mathrm{th}}^{L_0}$ &
 $+\Delta V_{\mathrm{th}}^{L_1}$ &
 $-\Delta V_{\mathrm{th}}^{L_0}$ &
 $0$ &
 $-\Delta V_{\mathrm{th}}^{L_0}$ \\
\addlinespace
$\mathbf{V_{\mathrm{REF2}}}$ &
 $+\Delta V_{\mathrm{th}}^{L_3}$ &
 $0$ &
 $0$ &
 $+\Delta V_{\mathrm{th}}^{L_3}$ &
 $0$ &
 $-(\Delta V_{\mathrm{th}}^{L_1}+\Delta V_{\mathrm{th}}^{L_2})$ &
 $+\Delta V_{\mathrm{th}}^{L_3}$ &
 $0$ &
 $0$ &
 $-\Delta V_{\mathrm{th}}^{L_2}$ \\
\addlinespace
$\mathbf{V_{\mathrm{REF1}}}$ &
$0$ &
$-\Delta V_{\mathrm{th}}^{L_1}$ &
$0$ &
 $0$ &
 $0$ &
 $0$ &
 $0$ &
 $0$ &
 $0$ &
 $0$ \\
\bottomrule
\end{tabularx}
\end{table}

\subsection{Practical implementation challenges}\label{sec:challenges}

Implementing MCFlash on COTS NAND flash memory chips, originally designed primarily for storage, presents several practical challenges. These challenges, however, are not fundamental barriers and can be resolved if NAND vendors provide enhanced flexibility in user-mode commands.

\textbf{Limited range for offset voltage:} Adjustable read offset voltages are primarily implemented in NAND flash memory to maintain data integrity and reliability in the presence of cell $V_{\mathrm{th}}$ variations due to wear or extended data retention. Although, theoretically, read offset voltages could extend across the entire memory window, in the chips under test, practical implementation constraints such as limited DAC bits often limit this range. Since typical operating conditions involve relatively minor threshold voltage shifts, such limited offset voltage ranges are enough to maintain data integrity. Consequently, the permissible offset voltage ranges vary significantly across vendors but generally remain narrow. Implementing MCFlash accurately requires voltage shifts to at least match the width of the target $V_{\mathrm{th}}$ distribution, $\Delta V_{\mathrm{th}}^{L_n}$, as depicted in Fig.~\ref{fig:4}. This requirement can be challenging for certain unmodified NAND chips, especially when attempting to shift $V_{\mathrm{REF0}}$ below the $L_0$ distribution. Note that NAND vendors typically provide a limited number of DAC (digital-to-analog converter) register values (e.g., 8-bit) for defining read reference offsets. Expanding this range would require minor parameter table extensions and potentially small updates in the DAC design to support a wider offset voltage span.
It is worth noting that most MCFlash operations (AND, OR, XNOR, NOT) run on commodity NAND chips without modification using user-mode commands. The other operations may show higher RBER (greater than \highlight{$5\%$)} due to limited read-offset ranges \highlight{in the chips under test}; expanding this range can be achieved using user-mode commands with additional \texttt{SET\_FEATURE} parameters. \highlight{Importantly, these limitations are not fundamental and can be addressed with modest extensions to vendor-provided read-offset configurations}. Thus, MCFlash remains fully compatible and deployable across different \highlight{generations} and technology nodes by only using user-based commands.

\textbf{Internal data randomizer and ECC:} 
Some raw NAND chips include built-in data randomizers and internal ECC engines, which can interfere with MCFlash operations unless these features are user-accessible and can be disabled via commands. However, most modern 3D NAND chips exclude such internal mechanisms, as data randomization and error correction are typically managed by the controller firmware or FTL layer. It is also worth noting that data-randomization-related reliability issues (wear-out) are inherent to NAND flash arrays and not specific to MCFlash; they similarly affect ParaBit and Flash-Cosmos.


\section{Experimental Evaluation}\label{sec:evaluation}

\subsection{Experimental setup and workflow}\label{sec:experimental-setup}

\begin{figure}[t]
\centering
\includegraphics[width=\linewidth]{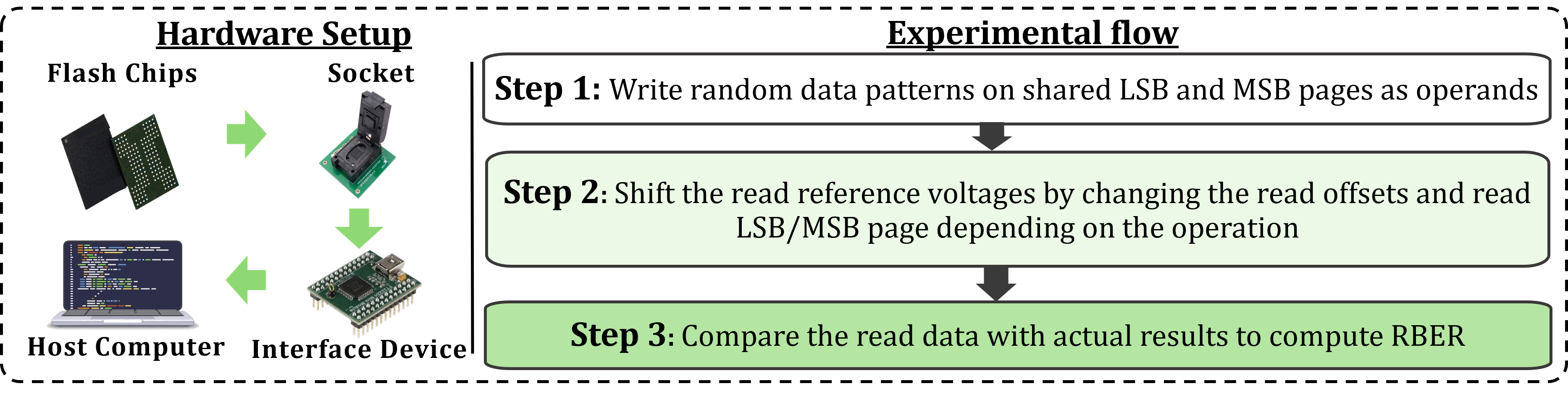}
\caption{Experimental setup and workflow for MCFlash implementation. A custom board with a BGA socket and USB interface provides direct access to raw NAND flash chips and ONFI-level commands.}
\label{fig:5}
\end{figure}

This study evaluates MCFlash on COTS 3D NAND flash memory chips from \highlight{a major vendors across three different generations}, incorporating two technology types: 176-layer charge-trap and 64-layer floating-gate as described in Table~\ref{tab:mcflash-rber}. A total of 50 individual chips were rigorously tested, comprising 10 chips from each product category and generation. To facilitate testing, a custom-built interface board was employed, as illustrated in Fig.~\ref{fig:5}. This specialized board integrates a Ball-Grid-Array (BGA) socket suitable for NAND flash packages and utilizes an FT2232H mini-module (Future Technology Devices International, FTDI)~\cite{ref42}, offering a USB connection to the host computer. Collectively, the FTDI module and host computer simulate the operations of a standard NAND flash controller.

Low-level firmware complemented by user-level wrapper functions were implemented to conform with the Open NAND Flash Interface (ONFI) standard command sets~\cite{ref43}. These software routines enabled low-level flash operations including page read, page program, and block erase, and supported advanced commands such as copyback~\cite{ref41}, read reference shifting, soft bit reading, and read retry via \texttt{SET\_FEATURE} user-mode commands. Critically, the experimental configuration facilitated direct access to the raw NAND flash array memory bits, deliberately bypassing any ECC and data randomization mechanisms. The comprehensive evaluation workflow for MCFlash implementation is depicted in Fig.~\ref{fig:5}.

From the SSD implementation perspective, MCFlash can be integrated at the firmware level of the FTL layer. In this design, the shared page data serves as operands, while the read offset sets for dedicated bitwise operations define the operation opcode. From the host side, each operation is invoked through dedicated APIs. The host interface layer schedules these operations independently from regular read operations and coordinates communication with the FTL layer.

For validation purposes, operand data were randomly generated and stored within logical shared-page buffers. A page programming operation subsequently transferred these shared pages to the selected memory layers. A strategically shifted read operation on the MSB/LSB pages, depending on the operation, then produced the desired bitwise operation results stored in the internal data buffer. This methodology permitted the repeated execution of various bitwise operations on the identical operand data through multiple shifted reads, without modifying the original data in the pages.

Reliability metrics, specifically the RBER in percentage, were calculated by systematically comparing actual outcomes against the expected results across different pages and blocks, ultimately deriving average failure rates. Consistency and reliability were further validated through cross-generation comparisons. Finally, measurements of read latency and energy consumption per operation were conducted using a Siglent SDS2354X oscilloscope and Nordic Power Profiler Kit II, enabling detailed performance analysis of timing and energy efficiency across diverse NAND devices.

\subsection{Successful implementation of MCFlash}\label{sec:mcflash-implementation}
The MCFlash technique was effectively realized on commercially available 3D NAND flash chips. Detailed performance evaluations of different MCFlash operations are summarized in Table~\ref{tab:mcflash-rber}. To guarantee robust statistical results, each bitwise operation underwent extensive testing involving a minimum of 1000 memory pages, each 16\,KB in size, translating into over one billion bitwise operations on 10 chips per part-number as shown in Table~\ref{tab:mcflash-rber}. 

\begin{table}[b]
\centering
\caption{Summary of MCFlash implementation results in terms of RBER for fresh and cycled blocks.}
\label{tab:mcflash-rber}
\setlength{\tabcolsep}{1.2pt}
\renewcommand{\arraystretch}{2.3}
\tiny
\begin{tabular}{ccc cccc cccc}
\hline
\textbf{Chip Count} &
\textbf{Part Number} &
\textbf{Part Description} &
\multicolumn{4}{c}{\textbf{Fresh [$N_{PE}=0$]}} &
\multicolumn{4}{c}{\textbf{Cycled [$N_{PE}=1.5$K]}} \\
 & & &
\textbf{AND} & \textbf{OR} & \textbf{XNOR} & \textbf{NOT} &
\textbf{AND} & \textbf{OR} & \textbf{XNOR} & \textbf{NOT} \\
\hline
10 & MT29F256G08EBHAFJ4 & 64-Layer FG &
0 & 0 & 0 & 0 &
0.00025\% & 0.000931\% & 0.00134\% & 0.00047\% \\

10 & MT29F512G08EEHAFJ4 & 64-Layer FG &
0 & 0 & 0 & 0 &
0.00019\% & 0.000846\% & 0.00124\% & 0.00032\% \\

10 & MT29F1T08EELEEJ4 & 176-Layer CT &
0 & 0 & 0 & 0 &
0.00012\% & 0.000763\% & 0.00108\% & 0.00069\% \\

10 & MT29F1T08EELKEJ4 & 176-Layer CT &
0 & 0 & 0 & 0 &
0.00009\% & 0.000821\% & 0.00119\% & 0.00057\% \\

10 & MT29F4T08GMLCEJ4 & 176-Layer CT &
0 & 0 & 0 & 0 &
0.00021\% & 0.000672\% & 0.00203\% & 0.00078\% \\
\hline
\end{tabular}
\end{table}

The implementation of two-operand AND, OR, XNOR and NOT demonstrated zero RBER under fresh conditions. The RBER for NAND, NOR and XNOR operations will be the same with inverse read. Note that this is the first demonstration of zero RBER for in-flash bulk bitwise operations across different COTS 3D NAND chips from \highlight{different technology generations.} \highlight{However, with increasing P/E cycle count, which reflects device endurance, non-zero RBER is observed, as shown in Table~\ref{tab:mcflash-rber}}. This trend aligns with anticipated outcomes stemming from the unavoidable threshold voltage distribution overlaps that intensify due to the broadening of distributions typical in MLC NAND storage under repeated P/E cycling.
\subsection{Influence of NAND flash cell reliability on MCFlash}\label{sec:cell-reliability}

\begin{figure}[b]
\centering
\includegraphics[width=\linewidth]{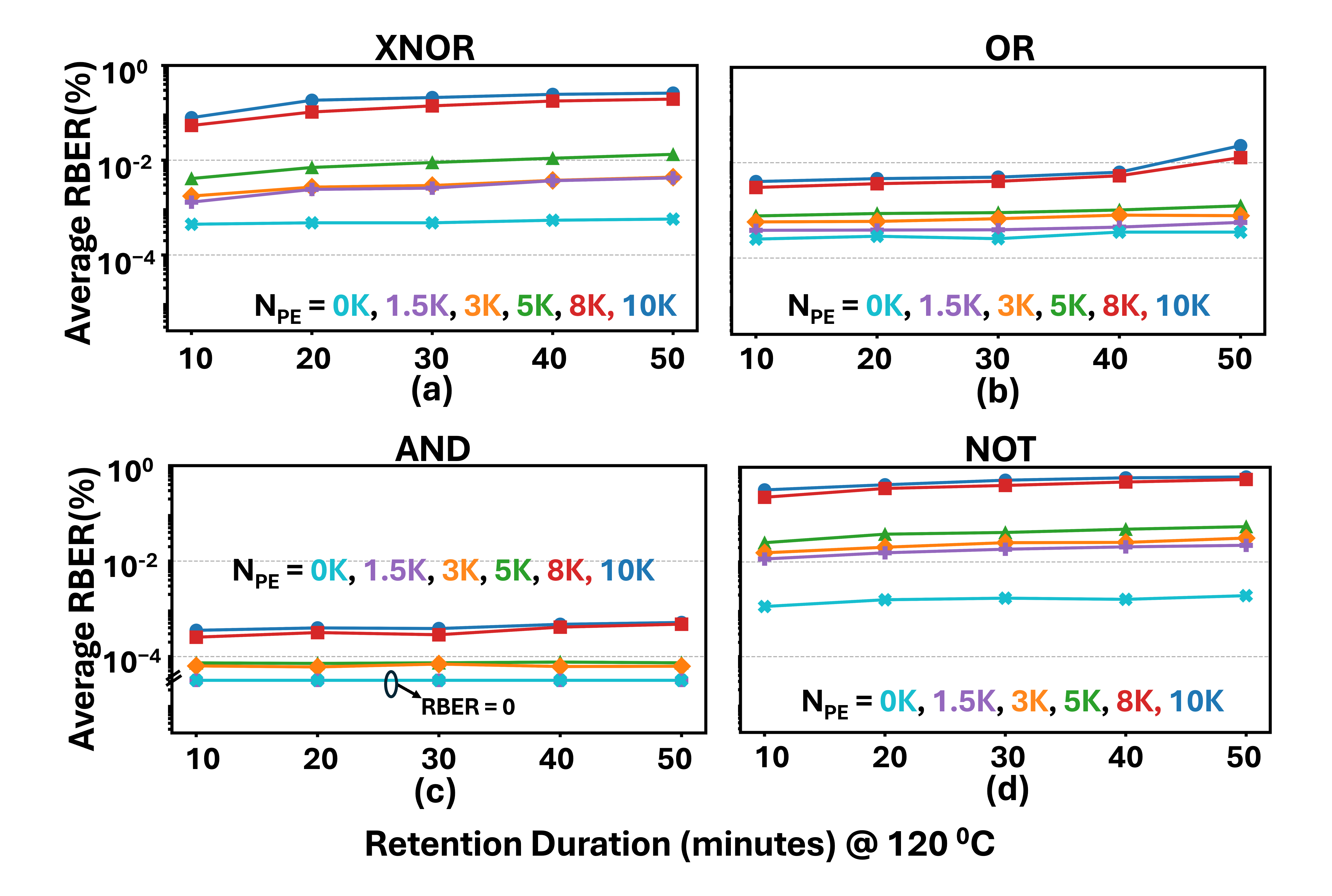}
\caption{Average RBER (\%) under increasing P/E cycles and retention duration at elevated temperature for (a) XNOR, (b) OR, (c) AND, (d) NOT}
\label{fig:6}
\end{figure}

This section investigates the impact of intrinsic NAND flash reliability such as endurance and retention characteristics on the accuracy of MCFlash operations. We assume that the operands are stored in the NAND array for a long duration before the bitwise operations are performed on them. Fig.~\ref{fig:6} presents experimental results illustrating the relationship between retention duration (x-axis) and RBER (y-axis) for bitwise OR, AND, XNOR and NOT operations across varying numbers of P/E cycles. Note that RBER is shown in log-scale for clear visualization, and hence the zero RBER points are separately shown. The experimental data clearly demonstrates a gradual increase in RBER as the NAND chips experience a higher number of P/E cycles and longer retention times, an expected outcome attributed to cell degradation effects.

Among the evaluated operations, the NOT and XNOR operation exhibits notably higher RBER than the other operations. This relatively higher error rate is due to the errors coming from the highest programmed state, $L_3$. Under retention duration, the $L_3$ state shifts the most, which adversely affects the NOT and the XNOR operations. Additionally, XNOR requires multiple internal sensing steps (typically four sensing operations), each of which independently contributes to the error probability. Errors from individual sensing operations add up, significantly increasing the cumulative RBER. Conversely, the AND operation consistently demonstrates a lower RBER. The reduced error rate for AND is primarily due to the simpler operational requirements involving only one sensing step and the favorable placement of the read reference voltage between the erase ($L_0$) and the lowest programmed state ($L_1$). Since the $L_1$ state shifts the least among all the programmed states, it reduces the probability of state misclassification, directly contributing to a lower RBER for the AND operation.

Indeed, the optimal read-offset value that minimizes RBER can depend on factors such as endurance, aging, and the physical location of memory cells. It is important to note, however, that the default read reference in commercial NAND arrays is already highly optimized by the manufacturer to minimize RBER under nominal conditions. This calibration inherently accounts for several sources of variability, including layer-to-layer differences, and in some devices, it is dynamically adjusted based on P/E cycling and aging characteristics of the block.

Since MCFlash operates by applying an additional offset on top of this default read reference, it inherits much of the robustness of the built-in calibration and is therefore not expected to be significantly impacted by intrinsic NAND variability. Nevertheless, the optimal offset value may still require fine-tuning to account for endurance degradation or long-term aging, and technology-specific characterization would be essential to ensure minimal RBER across all operating conditions.

In summary, while endurance degradation through P/E cycling and prolonged retention inevitably affect NAND flash reliability, thereby impacting the accuracy of MCFlash operations, judicious selection of sensing schemes, read reference voltages, and operational parameters can significantly mitigate these effects, maintaining a robust and acceptable error rate over extended use.

\subsection{Root-cause analysis and mitigation strategies of RBER}\label{sec:rber}

RBER is a critical metric for evaluating the reliability of MCFlash implementations in commercial NAND flash chips. In this section, we analyze the root causes of RBER in MCFlash and present potential mitigation strategies.

\begin{figure}[b]
\centering
\includegraphics[width=\linewidth]{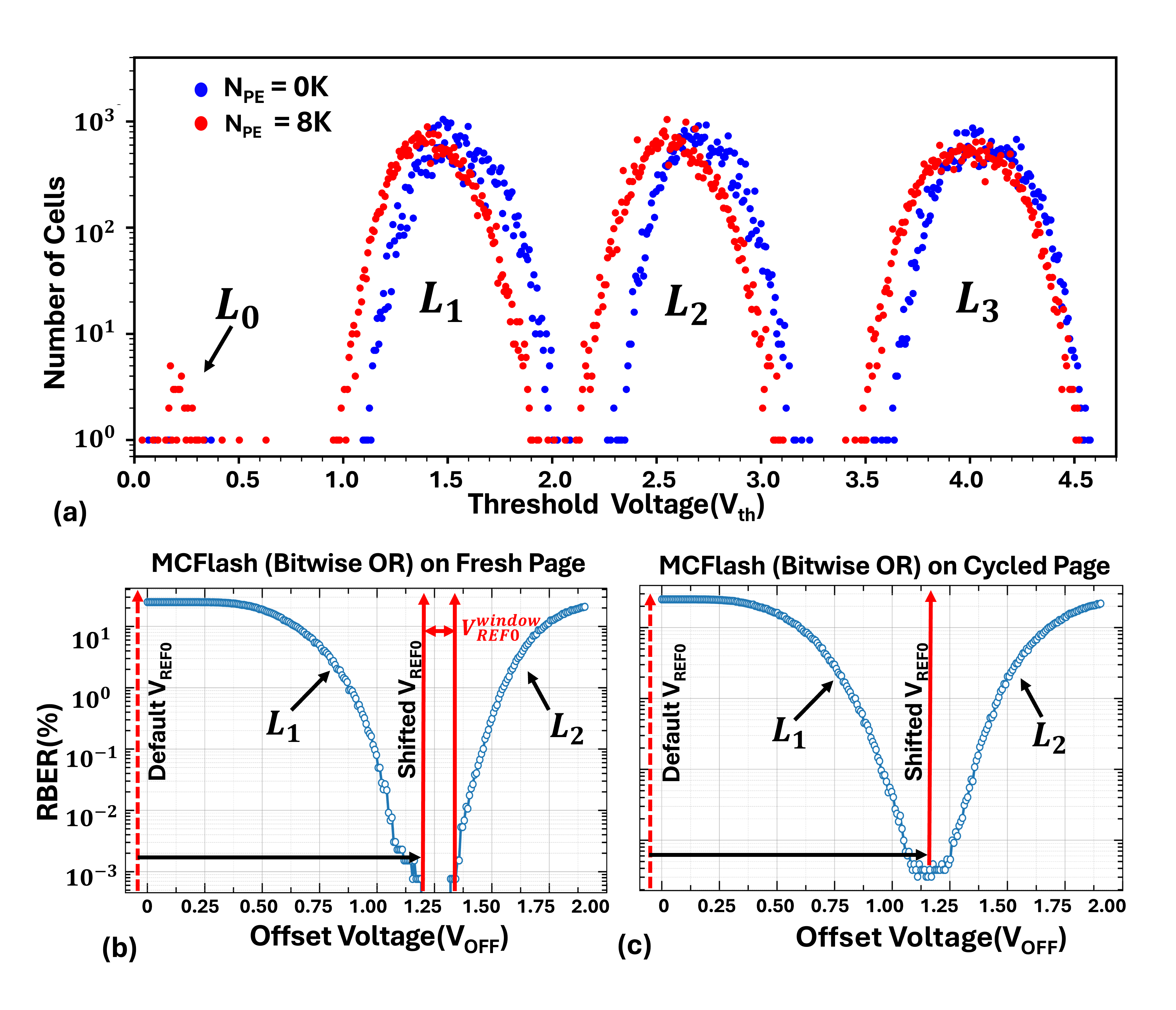}
\caption{Root-cause analysis of RBER in MCFlash. (a) Evolution of threshold voltage distributions with P/E cycling and resulting overlap between states. Impact of read-offset adjustment on RBER for bitwise OR in (b) fresh and (c) cycled blocks.}
\label{fig:7}
\end{figure}

Fig.~\ref{fig:7}(a) illustrates the underlying root-cause of RBER using the threshold voltage distributions of MLC storage. The blue dots represent $V_{\mathrm{th}}$ distributions on a fresh memory page, whereas the red dots represent distributions obtained on a cycled memory page with $N_{\mathrm{PE}} = 10\mathrm{k}$. We find that a narrow, non-overlapping voltage window exists between the two adjacent states for a fresh memory page. In this case, placing the shifted reference voltage within this window ensures correct bit classification, resulting in an RBER of zero. However, for the worn-out memory page ($N_{\mathrm{PE}} = 10\mathrm{k}$), the distribution widths are wider and shifted, causing diminishing voltage-window between adjacent states. As a result, even when the reference voltage is optimally placed at the valley between the two states, misclassification of bits near the overlap region occurs, leading to non-zero RBER. The distribution analysis in Fig.~\ref{fig:7}(a) highlights the importance of selecting appropriate read offset voltage, as it significantly influences the accuracy of MCFlash by minimizing corresponding RBER.

For example, as shown in Fig.~\ref{fig:4}(b), the bitwise OR operation requires appropriate placement of the shifted read reference voltage ($V_{\mathrm{REF0}}$) between $L_1$ and $L_2$ states. From Fig.~\ref{fig:7}(b), we observe that when the offset voltage ($V_{\mathrm{OFF}}$) is set to zero, the RBER is approximately \highlight{$25\%$}. This high error rate arises because all cells in state $L_1$ directly contribute errors. Since randomly generated operand data places about \highlight{$25\%$}
 of cells into state $L_1$, the RBER reflects the same percentage. As $V_{\mathrm{OFF}}$ is incrementally shifted upward, the RBER correspondingly decreases and eventually reaches zero once the offset surpasses the threshold voltage distribution width of $L_1$. Increasing $V_{\mathrm{OFF}}$ further within a certain range ($V_{\mathrm{REF0}}^{\mathrm{Window}}$) maintains an RBER of zero; however, excessive positive shifts beyond this window lead the read reference into the $L_2$ distribution, causing a subsequent increase in RBER due to misclassification errors.
Fig.~\ref{fig:7} highlights a critical observation: there exists an optimal voltage range for the offset voltage which ensures zero RBER for fresh NAND pages while it disappears for heavily cycled pages. With increasing P/E cycles, cells exhibit more unpredictable charge trapping and de-trapping behaviors. These phenomena broaden and shift the threshold voltage distributions for each programmed state, thereby elevating the likelihood of overlapping and shifting of the adjacent distributions and consequently increasing the inherent RBER. This trend is clearly illustrated in Fig.~\ref{fig:7}(c).

It is important to note that the optimal read-offset value that minimizes RBER can vary with factors such as endurance, aging, and the physical location of memory cells. However, commercial NAND flash chips are already factory-calibrated to minimize RBER under nominal conditions, with built-in mechanisms that account for variability such as layer-to-layer differences and, in some cases, dynamically adjust read references based on P/E cycling and aging. Since MCFlash applies an additional offset on top of this calibrated reference, it inherits much of this robustness and is unlikely to be significantly affected by intrinsic NAND variability.

Our examination of the spatial RBER distribution further supports that bit-flip locations within a memory layer are largely random and uniformly distributed, while systematic variations mainly appear across vertical layers in the 3D stack. These findings indicate that, although fine-tuning may still be required to accommodate endurance or long-term aging effects, MCFlash remains resilient to typical spatial and device-level variability.

\subsection{Latency and energy consumption of MCFlash}\label{sec:latency-energy}

The overall latency of MCFlash operations is primarily determined by the page read latency of the raw NAND flash chip. In MLC NAND, there are two types of pages: LSB and MSB, with different read times. LSB pages typically have lower latency (approximately $40\,\mu\mathrm{s}$) because they require only a single sensing phase using one reference voltage ($V_{\mathrm{REF1}}$ in Fig.~\ref{fig:2}). In contrast, MSB pages take longer to read (approximately $70\,\mu\mathrm{s}$) due to two sensing phases involving $V_{\mathrm{REF0}}$ and $V_{\mathrm{REF2}}$. Consequently, bitwise operations such as AND, which rely only on LSB page reads, incur lower latency than OR, NOT, or XNOR. Among these, XNOR has the highest latency, requiring up to four sensing phases when implemented using SBR techniques.

\begin{figure}[b]
\centering
\includegraphics[width=\linewidth]{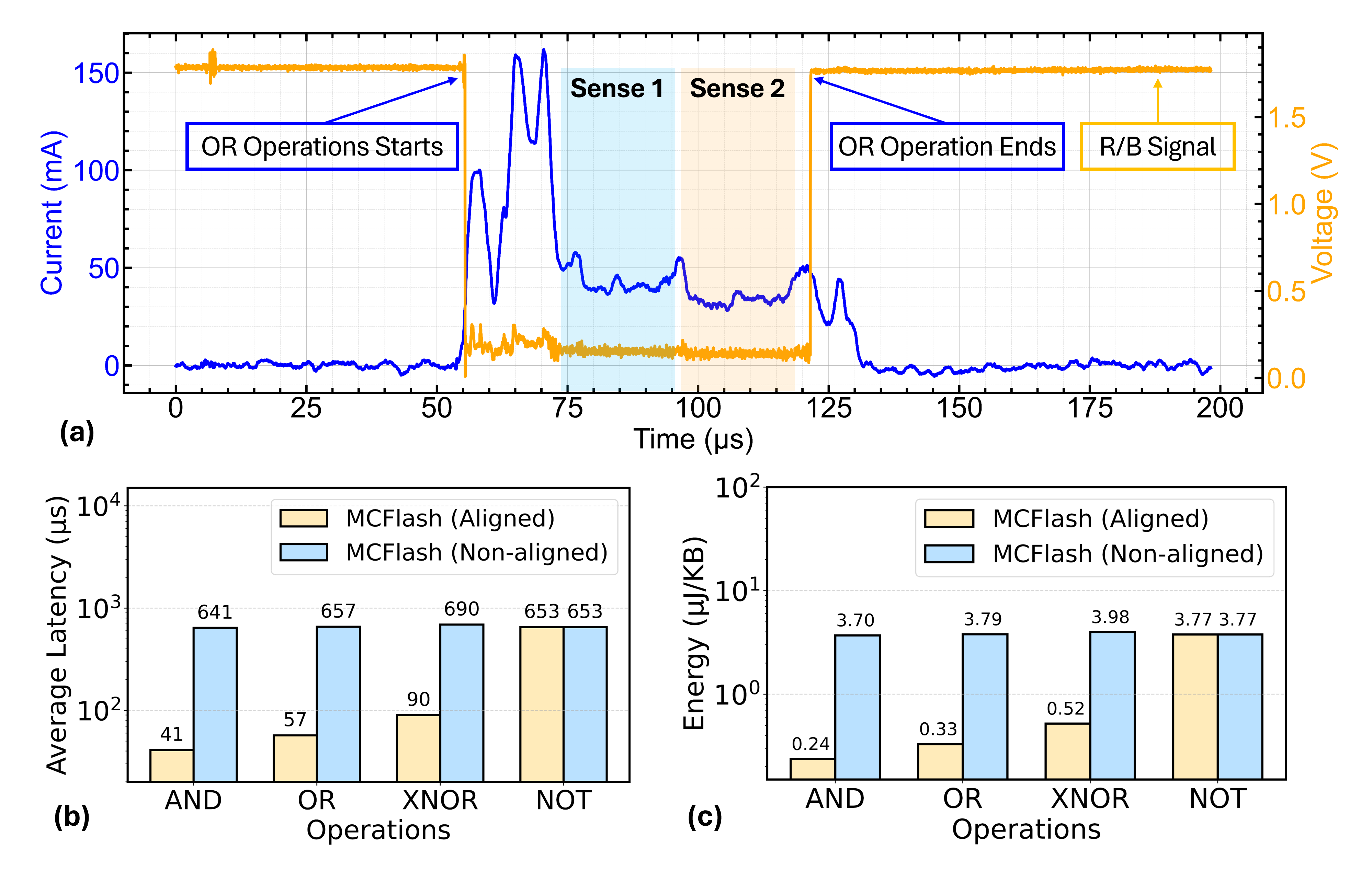}
\caption{Latency and energy characterization of MCFlash. (a) Measurement of the Ready/Busy (R/B) signal and supply current. (b) Latency comparison, (c) Energy consumption per kB for different bitwise operations.}
\label{fig:8}
\end{figure}

Fig.~\ref{fig:8}(a) illustrates our methodology for measuring the latency and energy consumption of page read operations in a raw NAND flash chip. We monitor the Ready/Busy (R/B) pin during the read process; the orange trace shows the digital status of the R/B signal. The period during which R/B remains low corresponds to the page read latency. Simultaneously, the blue trace captures the current drawn by the chip through its supply voltage pin, which is held at a constant voltage $V_{\mathrm{CC}} = 3.3\,\mathrm{V}$. For MSB pages, the current waveform shows an initial pre-charge spike followed by two distinct sensing intervals, reflecting the multi-phase sensing behavior. By integrating the current waveform over the page read interval, we measure the energy consumption corresponding to the page read operation, which also dictates MCFlash energy behavior.

Beyond the page read operation, MCFlash incurs a minor overhead (less than $10\,\mu\mathrm{s}$) for adjusting read reference offset voltages using “Set Feature’’ commands. This extra time when switching between different logic operations is minimal and has a negligible impact on overall MCFlash performance. Across all tested 3D NAND chips, the core page read latency ranges from approximately $40\,\mu\mathrm{s}$ to $90\,\mu\mathrm{s}$, with the sensing phase being the dominant contributor. As shown in Fig.~\ref{fig:8}(b), bitwise operations requiring fewer sensing phases, such as AND, which requires only a single sensing step, exhibit the lowest latency. In contrast, operations such as XNOR, which require up to four sensing phases, experience significantly higher latency.

This overhead can be greatly mitigated through internal SSD-level parallelism at the chip, die, and plane levels~\cite{ref44}. Fig.~\ref{fig:8}(b) also compares two MCFlash execution scenarios. The lower-latency case corresponds to aligned operands, where data resides on paired LSB and MSB pages along the same physical wordline. In contrast, when operands are not co-located, additional data alignment is required. This involves performing internal copyback operations using cache buffers within the flash die. We refer to this approach as MCFlash (non-aligned), illustrated later in Fig.~\ref{fig:9}(e). These extra program/read steps can extend the total per-page latency to approximately $600\,\mu\mathrm{s}$ to $800\,\mu\mathrm{s}$, depending on the NAND device generation. In this scenario, Flash-Cosmos supports multi-operand operations (up to 16), providing a significant advantage for executing long chains of bitwise operations compared to MCFlash and ParaBit.
Energy consumption was profiled at the chip level by partitioning each MCFlash operation into pre-charge, sensing, and discharge phases in Fig.~\ref{fig:8}(a). While the pre-charge and discharge energies remain largely invariant, the sensing energy scales linearly with the number of sensing phases. Fig.~\ref{fig:8}(c) quantifies this effect: per kilobyte, the XNOR operation consumes about \highlight{$51\%$} more energy than the AND operation. In scenarios requiring operand alignment, additional program/read cycles for copyback write further elevate energy usage, with the page program step dominating the incremental cost. These findings underscore that minimizing sensing phases and optimizing operand alignment are critical to maximizing both performance and energy efficiency in MCFlash-based logic operations on commercial NAND flash technology.

\subsection{Comparative analysis of energy consumption}\label{sec:energy-comparison}

\highlight{The energy consumption of IFC frameworks is primarily dictated by how bitwise operations are implemented (e.g., Flash-Cosmos, ParaBit, and MCFlash). In the following discussion, we consider only the computational energy, assuming aligned operands and negligible overhead from on-die NAND controller command execution.}

\highlight{For bitwise OR/NOR operations, Flash-Cosmos incurs higher energy consumption than MCFlash/ParaBit due to simultaneous activation of multiple blocks. For instance, a two-operand operation activates two blocks concurrently, resulting in approximately 34\% higher energy consumption than a standard page-read operation. As the number of operands increases, additional blocks must be activated, and when more than four blocks are involved, the power demand may exceed the budget of commodity SSDs~\cite{ref8}. In contrast, MCFlash performs OR/NOR operations without multi-block activation and therefore consumes energy comparable to a single page read. ParaBit exhibits similar energy efficiency, as both MCFlash and ParaBit rely on single-page, single-block sensing.}

\highlight{For AND/NAND operations, Flash-Cosmos shows comparable energy consumption to MCFlash for two-operand cases, since both require a single sensing cycle within a block. Flash-Cosmos, however, leverages MWS to perform bitwise operations across all wordlines in a block, which can offer advantages for multi-operand AND/NAND operations. That said, if MWS is accessible to users, MCFlash and ParaBit could similarly exploit it, resulting in comparable energy efficiency across all three frameworks.}

\highlight{For XOR/XNOR operations, Flash-Cosmos does not report detailed energy characterization. Both Flash-Cosmos and ParaBit employ inter-latch XOR logic, typically requiring six to eight sensing and latching steps~\cite{ref45, ref46}, leading to higher energy consumption. MCFlash instead utilizes SBR, as described in Section~\ref{sec:keymechanism}, requiring only two to four steps and thus achieving lower computational energy~\cite{ref39}.}

\highlight{If operands are not aligned, such as stored in different memory blocks, additional energy is required for operand realignment. A precise comparison in this case depends on framework-specific realignment mechanisms and is beyond the scope of this work. Overall, energy consumption across IFC frameworks varies significantly with the type of bitwise operation, and the energy efficiency of a given framework is therefore highly dependent on the application’s operation patterns, even when considering computational energy alone.}


\section{System-Level Evaluation: MCFlash and Alternatives}\label{sec:system-eval}

Fig.~\ref{fig:9}(a) illustrates the architecture of the target SSD used for the system-level performance evaluation of MCFlash. The SSD consists of 16 channels, each connected to eight 4-plane NAND dies, resulting in a total of 512 planes. Each plane supports a page size of 16\,kB. The channel-to-controller bandwidth is assumed to be 1.2\,GB/s, while the interface between the host and the SSD is modeled as a 4-lane PCIe Gen4 link with a total bandwidth of 8\,GB/s. For evaluation, we assume that bulk bitwise operations are performed on 8\,MB-long bit vectors, which are evenly distributed across all 512 planes, as depicted in Fig.~\ref{fig:9}(a). To estimate the highest performance, we assume that the host can issue concurrent multi-plane reads across all NAND dies for each operand, representing a best-case scenario for maximizing throughput.

\subsection{Comparison of MCFlash and alternatives}\label{sec:comparison}

\begin{figure}[b]
\centering
\includegraphics[width=\linewidth]{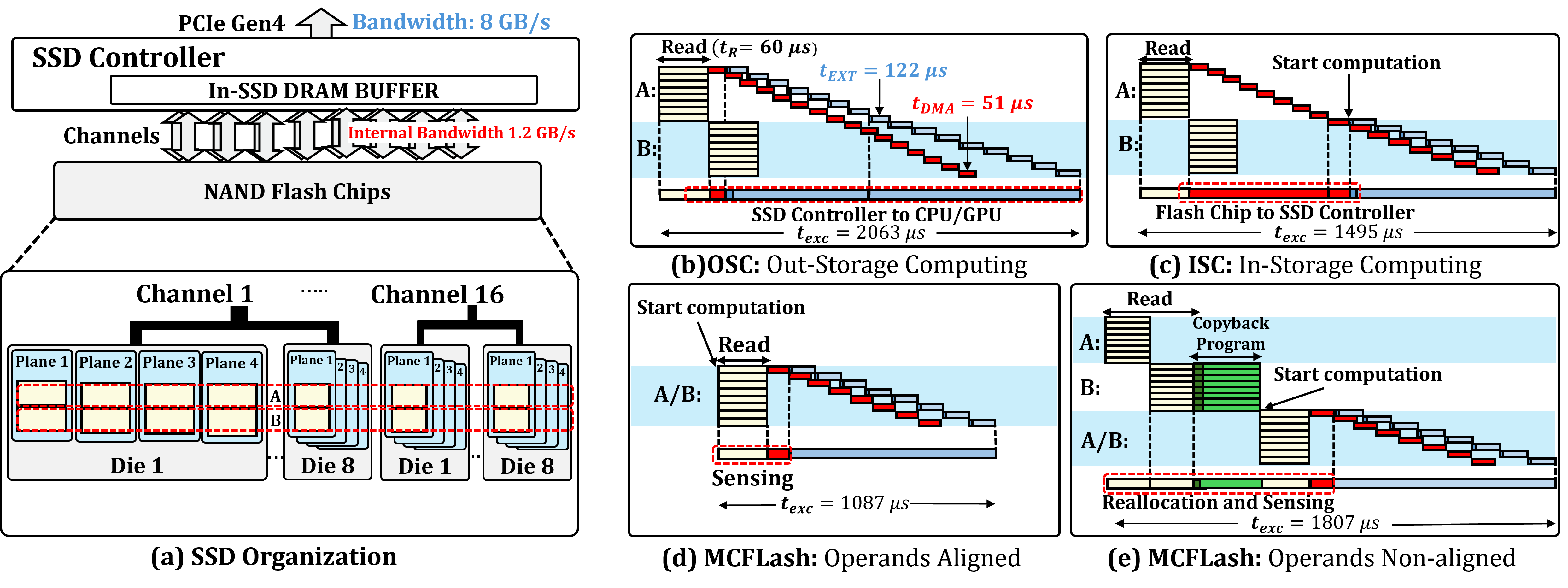}
\caption{System-level execution of bulk bitwise operations. (a) Target SSD organization. (b) Timeline for outside-storage computing (OSC). (c) In-storage computing (ISC). (d) MCFlash with aligned operands. (e) MCFlash with non-aligned operands requiring operand alignment.}
\label{fig:9}
\end{figure}

Fig.~\ref{fig:9}(b) shows the execution timeline for a channel when an application uses the traditional outside-storage computing (OSC) paradigm to perform bulk bitwise operations on two 8\,MB bit vectors $A$ and $B$. The operands are read sequentially as illustrated in the figure. After an operand is loaded into the sensing latch, it can be transferred to the SSD controller and then forwarded to the host, while the next operand is being read in parallel. Given the internal flash-channel bandwidth, the time required for each die to transfer data from the sensing latch to the controller is $t_{\mathrm{DMA}} = \frac{4 \times 16\,\text{kB}}{1.2\,\text{GB/s}} \approx 51\,\mu\text{s}$. Since all channels transfer data concurrently to the controller, after $t_{\mathrm{DMA}} = 51\,\mu\text{s}$ the controller will hold 1\,MB ($16\,\text{kB} \times 4 \times 16$) of data that must be transferred serially to the host, which will take $t_{\mathrm{EXT}} = 16 \times \frac{4 \times 16\,\text{kB}}{8\,\text{GB/s}} \approx 122\,\mu\text{s}$. Thus, the overall execution time for this example is $t_{\mathrm{exc}}^{\mathrm{OSC}} = t_R + t_{\mathrm{DMA}} + 16t_{\mathrm{EXT}} = 2063\,\mu\text{s}$, assuming $t_R = 60\,\mu\text{s}$. Clearly, performance is constrained by the external I/O bandwidth. As shown in Fig.~\ref{fig:9}(b), the SSD must sequentially transfer all bit vectors from its 16 channels through the host interface, creating a bottleneck when computations are performed on the host CPU.
Fig.~\ref{fig:9}(c) illustrates the execution timeline for a single channel under the ISC paradigm, where bitwise operations are performed within the SSD controller and only the final result is sent to the host. In this scenario, the internal data transfer between the NAND flash dies and the SSD controller becomes a major contributor to overall latency. The total execution time is calculated as $t_{\mathrm{exc}}^{\mathrm{ISC}} = t_R + 9t_{\mathrm{DMA}} + 8t_{\mathrm{EXT}} = 1495\,\mu\text{s}$. Compared to the OSC case, ISC provides a notable improvement in the overall execution time.

Fig.~\ref{fig:9}(d) shows the execution timeline for a single channel for the MCFlash approach. Here we assume that both operands are properly aligned, meaning they are stored in the corresponding shared pages (LSB and MSB sharing the same wordline) across the dies. Thus, a single read operation computes the bitwise operation result, which is then transferred to the SSD controller and the host. The total execution time is calculated as $t_{\mathrm{exc}}^{\mathrm{IFC}} = t_R + t_{\mathrm{DMA}} + 8t_{\mathrm{EXT}} = 1087\,\mu\text{s}$. Note that the page read time ($t_R$) depends on the type of bitwise operation. For instance, a bitwise AND can be performed using an LSB page read, which is faster, while a bitwise OR may require reading the MSB page, which takes comparatively more time.

Finally, Fig.~\ref{fig:9}(e) shows the execution timeline for the MCFlash approach when operand alignment is required before computation. If this alignment must be done at runtime, it introduces significant delays to the overall execution time. For example, runtime alignment requires reading operands from their original (and possibly scattered) pages and writing them to specific MLC shared pages located on the same wordline. This step can leverage the internal copyback write mechanism, which uses the die’s internal buffer to move data without incurring external data transfer latency.

Although copyback avoids host-side transfers, it still adds substantial delay because the MLC page write time ($t_{\mathrm{prog}} \approx 600\,\mu\text{s}$) is much longer than read time. As a result, the total execution time becomes $t_{\mathrm{exc}}^{\mathrm{IFC}}(\text{non-aligned}) = 3t_R + t_{\mathrm{prog}} + t_{\mathrm{DMA}} + 8t_{\mathrm{EXT}} = 1807\,\mu\text{s}$. This overhead can be avoided if operands are pre-aligned ahead of time, for example through workload profiling or prediction of computation patterns.

\subsection{Impact on application-level case studies}\label{sec:app-impact}

We evaluated the application-level impact of MCFlash by comparing its performance with an outside storage computing system (OSC), an in-storage computing system (ISC),\highlight{ ParaBit~\cite{ref7} and FlashCosmos~\cite{ref8}}. The configurations for OSC, ISC, ParaBit and FlashCosmos are adopted directly from the evaluation setup in \highlight{~\cite{ref8}}. The SSD configuration we considered for the estimation consists of 16 channels, 8 dies per channel and 4 planes per die. The size of each flash page is 16\,KB.

\textbf{Image Segmentation:} We use an image segmentation workload based on YUV color recognition, where each pixel is classified into one of four predefined color classes using bitwise AND operations. The workload operates on images of resolution $800 \times 600$ and is evaluated across dataset sizes ranging from 10K to 200K images. For each pixel $m$, the color recognition result is expressed as
$\mathrm{Re}(m) =
(C_{1,a}, C_{1,b}, C_{1,c}, C_{1,d})
\ \mathrm{AND}\
(C_{2,a}, C_{2,b}, C_{2,c}, C_{2,d})
\ \mathrm{AND}\
(C_{3,a}, C_{3,b}, C_{3,c}, C_{3,d})$,
where $C_1$, $C_2$ and $C_3$ correspond to the Y, U and V channels respectively, and $a,b,c,d$ denote the four target classes. On average, MCFlash achieves $16.5\times$, $12.69\times$, $1.76\times$ and $0.5\times$ speedup over OSC, ISC, ParaBit and FlashCosmos, respectively. 

\textbf{Image Encryption:} We further evaluate MCFlash on image encryption workloads, which consist of bulk bitwise XOR operations. Each bit of every pixel in an RGB image is encrypted by performing a bitwise XOR with a key of the same size. The workload size varies from 5K to 100K images. The encryption process is defined as
$\mathrm{Cipher}(x) = \mathrm{Image}(x) \oplus \mathrm{Key}(x)$,
where $x$ denotes the pixel location. On average, MCFlash achieves $20.92\times$, $16.02\times$, $2.22\times$ and $0.63\times$ speedup over OSC, ISC, ParaBit and FlashCosmos, respectively. 

\textbf{Bitmap Indices:} We also evaluated MCFlash using a bitmap index workload, which is commonly used in databases to address analytical queries. We assumed there are 800 million users and aimed to compute the number of users who were active every day over $m$ months. For each user $y$, this value is calculated as
$\mathrm{Res}(y) = (V_{1,y}) \ \mathrm{AND}\ (V_{2,y}) \ \mathrm{AND}\ \ldots\ \mathrm{AND}\ (V_{x,y})$,
where $x$ denotes the number of days, and $V_{x,y}$ is a bit that indicates whether user $y$ was active on day $x$. This computation involves a chain of bitwise AND operations, which can be executed on MCFlash, and a bit-count operation, which is offloaded to the processor. The evaluation time varied between 1 to 12 months. On average, MCFlash achieves $31.67\times$, $24.26\times$, $3.37\times$ and $0.96\times$ speedup over OSC, ISC, ParaBit and FlashCosmos, respectively.

Across the applications, MCFlash offers significant performance gains for real-time, bitwise-intensive applications by exploiting a combination of MLC sensing optimizations and high page-level parallelism. Unlike conventional OSC systems that suffer from significant data movement overhead between storage and memory, MCFlash eliminates host-controller bottlenecks by performing bitwise operations directly within flash arrays. Compared to ISC, MCFlash avoids both memory-controller transfers and internal SSD bandwidth limitations by leveraging fully in-memory computation. ParaBit uses the external DRAM buffer inside SSD for reallocation, and this adds a latency overhead. On the other hand, MCFlash exploits the user command copyback, which utilizes the internal cache register of the NAND flash chip, avoiding external data movement. FlashCosmos can execute multi-operand (more than two) bitwise operations in one sensing cycle by leveraging multi-wordline and multi-block sensing operations within the NAND flash array. However, as MWS is not a commodity user command, FlashCosmos lags in technology readiness. MCFlash is deployable on commercially available flash chips without hardware modifications and leverages user commands only. MCFlash's latency scales linearly with workload size, confirming predictable performance and efficient utilization of all flash chips. These advantages make MCFlash a scalable and high-performance solution for in-storage acceleration of real-time, data-intensive workloads.

\begin{figure}[t]
\centering
\includegraphics[width=\linewidth]{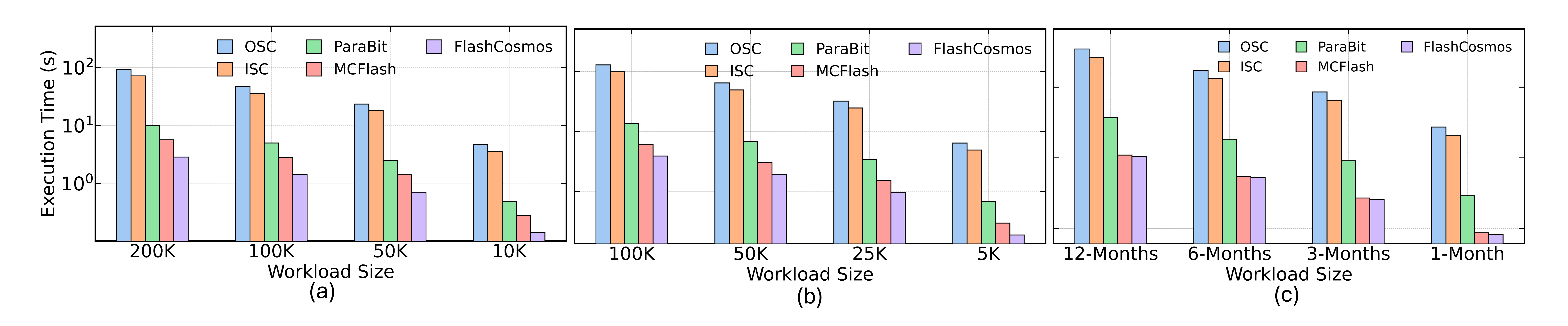}
\caption{Execution time of different architectures with respect to MCFlash for three application workloads: (a) Image segmentation, (b) Image encryption, and (c) Bitmap indices.}
\label{fig:10}
\end{figure}


\section{Discussion}\label{sec:discussion}

\highlight{Although our discussion of MCFlash focuses on two-operand bitwise operations, the same principle supports three-operand operations in Tri-Level Cell (TLC) and four-operand operations in Quad-Level Cell (QLC) memory. These extensions rely on the additional shared pages due to logical scaling provided by higher-density NAND technologies.} Moreover, TLC devices can be operated in a reduced-MLC mode by initializing one of the shared pages with a fixed data pattern, such as all-ones or all-zeros~\cite{ref47}. This approach allows more robust sensing with improved threshold voltage spacing, potentially enabling zero RBER even for worn-out memory and enhancing overall reliability.

While MCFlash may not scale to very large multi-input operations as extensively as Flash-Cosmos, this trade-off reflects its focus on compatibility with commodity NAND flash chips and reliance on firmware-level modifications only. MCFlash maintains high reliability and efficiency for bitwise and small multi-input operations, while larger logic functions can still be executed through sequential operand composition with modest latency overhead.

Second, similar to ParaBit and Flash-Cosmos, MCFlash enables efficient bulk bitwise operations only when the participating operands reside within the same NAND chip. To support such locality, the system could implement background data alignment mechanisms, such as inter-chip operand migration to co-locate data on shared wordlines. However, these migrations introduce additional data movement overheads, which can diminish the performance gains offered by IFC. In scenarios where the required operands are distributed across multiple chips, alternative architectures, such as ISC solutions that employ dedicated hardware accelerators near the flash interface, may offer superior efficiency by reducing inter-chip coordination and movement costs.

Third, Cross-layer error mitigation strategies can be effectively applied to MCFlash to achieve zero RBER. To begin with, the $V_{\mathrm{th}}$ distribution width of MLC cells can be reduced by adopting a smaller program step size in the ISPP algorithm, which, although increasing programming latency, significantly lowers RBER. In addition, the read-offset values can be dynamically optimized based on cell state, spatial location, and aging conditions, thereby minimizing RBER at the expense of added calibration complexity. Furthermore, TLC devices can operate in a reduced MLC mode by utilizing only four of the eight $V_{\mathrm{th}}$ states, which enlarges the voltage margin between states; however, this approach may still face endurance limitations due to the inherently lower durability of TLC cells.

Finally, MCFlash performance can be further enhanced with vendor-level support. For instance, expanding the permissible range of read-offset values, particularly to allow shifting the reference voltage below the erase distribution ($L_0$), would simplify on-chip implementations of operations like NAND, NOR, and XOR, enabling more efficient and accurate execution within the flash array.


\section{Conclusion}\label{sec:conclusion}

In this paper, we introduced MCFlash, a highly practical and immediately deployable technique enabling bulk bitwise operations directly within COTS 3D NAND flash memory chips, without requiring any hardware modifications or privileged command sequences. MCFlash leverages dynamic read reference voltage shifting combined with a specialized logical data encoding strategy to perform efficient and accurate two-operand bulk bitwise operations. Experimental results demonstrate that MCFlash reliably achieves zero RBER for NOT, OR, AND, and XNOR operations on fresh memory pages and sustains an impressively low RBER below 0.015\% even after P/E cycling. Additionally, our application-level evaluation on image segmentation, encryption, and bitmap indices revealed significant performance gains, with MCFlash outperforming ParaBit by up to 1.7$\times$, and exceeding OSC and ISC accelerator implementations by more than 16.5$\times$ and 12.6$\times$ respectively.

Overall, MCFlash offers an effective and reliable solution for in-flash computation on commercial NAND devices. Future work will focus on system-level integration, extending supported operations via advanced voltage encoding, and evaluating performance across broader workloads.

\bibliographystyle{unsrt}
\bibliography{sn-bibliography}

\end{document}